\documentclass[preprint,superscriptaddress,amsmath,amssymb,floatfix,prb]{revtex4-2}
\usepackage{amsmath,amssymb,mathtools}
\usepackage{graphicx}
\usepackage{bm}
\usepackage[colorlinks, citecolor=blue, linkcolor=red]{hyperref}
\usepackage{braket}
\usepackage{csquotes}
\usepackage{hyperref}
\usepackage{xcolor, color, soul}

\begin{document}

\preprint{APS/123-QED}

\title{Bipartite entanglement via distance between the states in a one dimensional spin 1/2 dimer copper acetate monohydrate}

\author{S. Athira}
\affiliation{School of Physics, IISER Thiruvananthapuram, Vithura, Thiruvananthapuram-695551, India}
\author{Saulo L. L. Silva}
\affiliation{Centro Federal de Educação Tecnológica, CEFET-MG, Nepomuceno-37250000, Brazil}
\author{Sushma Lakshmi}
\affiliation{School of Physics, IISER Thiruvananthapuram, Vithura, Thiruvananthapuram-695551, India}
\author{Sharath Kumar C}
\affiliation{School of Physics, IISER Thiruvananthapuram, Vithura, Thiruvananthapuram-695551, India}
\author{Debendra Prasad Panda}
\affiliation{School of Advanced Materials, and Chemistry and Physics of Materials Unit, Jawaharlal Nehru Centre for Advanced Scientific Research, Jakkur, Bangalore-560064, India}
\author{Sayan Das}
\affiliation{School of Chemistry, IISER Thiruvananthapuram, Vithura, Thiruvananthapuram-695551, India}
\author{Probal Nag}
\affiliation{School of Chemistry, IISER Thiruvananthapuram, Vithura, Thiruvananthapuram-695551, India}
\author{Andrews P. Alex}
\affiliation{School of Physics, IISER Thiruvananthapuram, Vithura, Thiruvananthapuram-695551, India}
\author{A. Sundaresan}
\affiliation{School of Advanced Materials, and Chemistry and Physics of Materials Unit, Jawaharlal Nehru Centre for Advanced Scientific Research, Jakkur, Bangalore-560064, India}
\author{Sivaranjana Reddy Vennapusa}
\affiliation{School of Chemistry, IISER Thiruvananthapuram, Vithura, Thiruvananthapuram-695551, India}
\author{D. Jaiswal-Nagar}
\email{deepshikha@iisertvm.ac.in}
\affiliation{School of Physics, IISER Thiruvananthapuram, Vithura, Thiruvananthapuram-695551, India}

\date{\today}

\begin{abstract}
In this paper, we used a theoretical measure known as distance between the states, $\mathcal{E}(\rho_e)$, to determine the bipartite entanglement of a one dimensional magnetic dimer system. The calculation was compared with the well-known entanglement measure, concurrence, and found to be the same. $\mathcal{E}(\rho_e)$ was, then, expressed in terms of two thermodynamic quantities, namely, magnetic susceptibility and specific heat. Experimental verification of temperature variation of the bipartite entanglement measure in terms of magnetic susceptibility and specific heat was done on single crystals of copper acetate-an excellent one dimensional dimer system. The results showed the existence of bipartite entanglement till temperatures as high as room temperature! Large sized single crystals of copper acetate were grown by a new evaporation technique and characterised by TGA, IR and Raman spectroscopy measurements.Density functional theory calculations were done to calculate the delocalisation index which showed much lower values of $\delta(Cu,Cu)$ than other bonds, implying that the probability of direct Cu-Cu exchange in copper acetate is very small. 
\end{abstract}

\maketitle


\section{\label{sec:level1}Introduction\protect\\ }

Low dimensional magnets have been at the forefront of research recently due to pronounced quantum effects seen in them at finite temperatures as a consequence of low dimensions \cite{duffy}. The quantum effects are expected to be more pronounced when the total spin quantum number S of the low dimensional magnet is small: smaller the spin quantum number, higher the quantum fluctuations\cite{Pires}. So, a one dimensional quantum magnet with S = 1/2 is expected to show the most pronounced quantum mechanical effects since the spatial dimensionality is the lowest at 1 and the spin quantum number S is lowest at 1/2. Therefore, spin 1/2 ions such as Cu(II) have been extensively used to study low dimensional quantum molecular magnetism. In recent times, low dimensional magnets have also been studied extensively in the area of quantum information processing where the S = 1/2 spin chains are treated as a bus that link quantum information processors \cite{bose, apollaro, nikolopolos}. Here, entanglement between spins results in superposition states of spins being held for a long time and consequently, entanglement being used as a resource. Entanglement  exemplified as a superposition of two spin $1/2$ states, ($1/\sqrt{2}$)($\ket{\uparrow\downarrow}$ - $\ket{\downarrow\uparrow}$) is one of the most intriguing quantum mechanical phenomena where a coherent superposition of states cannot be written as product states of individual wavefunctions \cite{wootters,conner,arnesen,wang,gu,vedral}. Using such entangled states, quantum protocols such as teleportation \cite{ekert,deutsch,nielsen}, can be realised for perfect transportation of quantum states. There exist various measures to quantify entanglement \cite{hao,venuti,bose,wootters,conner,gu,vedral,amico,guhne,
osborne,george}, however, no unique measure of entanglement is expected even for mixed bipartite states \cite{conner,vedral,amico}.\\ 
Spin dimer Heisenberg antiferromagnetic chain (HAfc) systems are those systems in which the nearest neighbour spins interact with each other via two exchange coupling constants J$_1$ and $J_2$ such that J$_2 <<$ J$_1$ resulting in pair of spins being alternately coupled by strong and weak bonds governed by two different values of exchange coupling constants \cite{bleaney,ghoshray,urushihara}. The ground state of a dimer HAfc system is non-magnetic comprising spin singlets and is separated from a triplet excited state (consisting of triplons) by a finite spin gap $\Delta$. When J$_1$ is not appreciably different from J$_2$, the dimeric chain is called an alternating chain defined with an alteration parameter $\alpha$ = J$_2/$J$_1$, such that 0 $ < \alpha <$ 1. The alternate chain systems with variable values of $\alpha$ have been in intense limelight recently due to their relevance in quantum information theory, wherein, they have been used as a model system in studying bipartite entanglement measures like concurrence \cite{conner, arnesen,brukner}.\\
A finite interdimer interaction converts the dimer system to an alternating chain material affecting the value of the energy gap $\Delta$. In order for a spin dimer to be used as an information storage or a quantum computing material, it should mimic the idealiased dimer system as closely as possible. The various reported spin dimer systems have a value of $\alpha$ that is small but in the range of 0.1-0.3 \cite{spodine,Bonner,Xu,He,nawa,urushihara}. However, the celebrated dimer system tetrakis(acetate) diaquadicopper (C$_8$H$_{16}$Cu$_2$O$_{10}$)-copper acetate for short \cite{Koziskova} has a very small value of the alteration parameter $\alpha$ and represents the model system on which entanglement measures for a dimer system could be tested. 
In this paper, we employ a bipartite measure of entanglement, namely, "distance between the states" for a magnetic dimer system in terms of which two thermodynamic quantities, namely, magnetic susceptibility and specific heat is expressed. The equivalence of this measure to a well-known measure, namely, concurrence is also shown. The calculations are tested on copper acetate-the celebrated one dimensional dimer system, that was grown using a new synthesis method resulting in large-sized single crystals. The structure of copper acetate was confirmed with single crystal x-day diffraction. Fourier transform infrared spectroscopy and Raman spectroscopy was done to identiy the prominent relevant bands in the structure that were confirmed by density functional theory calculations employing a mixed basis having a metal centre and an organic part \cite{georgeCRAT, sharath}. Further, delocalisation index was calculated for different bonds and it was found that the delocalisation index value for direct Cu-Cu bonds was much lower than other bonds implying that the probability of direct Cu-Cu exchange in copper acetate is quite small.  

\section{Theory}

\subsection{Bipartite Entanglement measure: Distance between the states}

Vedral et al. \cite{vedraldistance} proposed a new measure of entanglement, namely, distance between the states, $\mathcal{D}$, in the quantum mechanical state space, where the distance is not necessarily a distance in the metric sense. Consider a set of states represented by $\Omega$ density matrices in a 2 $\otimes$ 2 Hilbert space such that $\Omega$ contains a subset of separable states, $\mathcal{S}$, and a subset of entangled states, $\Sigma = \Omega - \mathcal{S}$. The entanglement measure $\mathcal{E}(\rho_e)$ of a state $\rho_e \in \Sigma$  is defined as \cite{vedraldistance}: 
\begin{equation}
\mathcal{E}(\rho_e) = \min_{\rho_s \in \mathcal{S}}\mathcal{D}(\rho_e, \rho_s)
\label{eqn:vedral}
\end{equation}
where $\rho_e$ and $\rho_s$ denote the density matrix associated with entangled and separable states respectively.\\
Out of the several possibilities that could be used to define such a distance $\mathcal{D}$, Witte et al. \cite{witte} used the Hilbert-Schimdt norm as a measure of the distance $\mathcal{D}$ to calculate the entanglement measure $\mathcal{E}(\rho_e)$ as:
\begin{equation}
\mathcal{E}(\rho_e) = \min_{\rho_s \in \mathcal{S}}||\rho_s - \rho_e||^2
\label{eqn:H-S norm}
\end{equation}
Witte et al. \cite{witte} also showed that $\mathcal{E}(\rho_e)$ is a good measure of entanglement satisfying the various requirements of an entanglement measure.\\  
Using the Hilbert-Schimdt norm, del Cima et al. \cite{delcima} showed the entanglement measure to be:
\begin{equation}
\mathcal{E}(\rho_e)=2\epsilon_0\mathrm{max}(0,\vert z \vert -v)
\label{eqn:distance}
\end{equation}
where z and $\nu$ are the matrix element of reduced density matrix $\rho_e$ and $\rho_s$ and $\epsilon_0$ is a normalization constant to ensure that the entanglement measure $\mathcal{E}(\rho_e)$ satisfies:
$$0 \leq \mathcal{E}(\rho_e) \leq 1$$.
and In order to calculate this measure for a dimer system, we consider the Hamiltonian for a spin 1/2 Heisenberg antiferromagnetic dimer chain that is given by:
\begin{equation}
H = -J\sum\limits_{i=1}^{N/2}(\overrightarrow{S}_{i-1} \cdot \overrightarrow{S}_{i} + \alpha \overrightarrow{S}_{i} \cdot \overrightarrow{S}_{i+1}) 
\label{eqn:Hamiltonianalpha}
\end{equation}
where $J < 0$ represents the exchange coupling constant between a spin at the $i^{th}$ site and its nearest neighbour spin at the $(i-1)^{th}$ site while $\alpha J$ represents the exchange interaction between the same $i^{th}$ spin but with the other nearest $(i+1)^{th}$ neighbour.  $\overrightarrow{S}_{i-1}$, $\overrightarrow{S}_{i}$ and $\overrightarrow{S}_{i+1}$ represent the spin operators on the $(i-1)^{th}$, $i^{th}$ and $(i+1)^{th}$ lattice sites respectively. $N$ denotes the number of sites in the dimer. In terms of $\alpha$, the energy gap is defined as \cite{johnston}:
\begin{equation}
\Delta = J_1(1-\alpha)^{\frac{3}{4}}(1 + \alpha)^\frac{1}{4} 
\label{eqn:delta}
\end{equation}
where $0\le\alpha\le 1$.\\
For a dimer, $\alpha$ = 0. Considering a bipartite system where one sub-system is a dimer and the second sub-system is the remaining lattice, equation \ref{eqn:Hamiltonianalpha} can be rewritten as:
\begin{equation}
H = -J\vec{S}_1 \cdot \vec{S}_2
\label{eqn:Hamiltoniandimer}
\end{equation}

In terms of the spin matrices $S^x$, $S^y$ and $S^z$, equation \ref{eqn:Hamiltoniandimer} becomes:
\begin{equation}
H=-J\left(S^{x}_{1}S^{x}_{2}+S^{y}_{1}S^{y}_{2}+S^{z}_{1}S^{z}_{2} \right)
\label{eqn:HamiltSx}
\end{equation}

Moreover, $S^{\alpha}_{i}=\frac{1}{2}\sigma^\alpha$ where $\sigma$ is the Pauli's matrix. Considering $\hbar=1$, equation \ref{eqn:HamiltSx} can be written in the matrix form as:

\begin{equation}
H=\left(\begin{array}{cccc}
-\frac{J}{4} & 0 & 0 & 0\\
0 & \frac{J}{4} & -\frac{J}{2} & 0\\
0 & -\frac{J}{2} & \frac{J}{4} & 0\\
0 & 0 & 0 & -\frac{J}{4}
\end{array}\right),
\end{equation}
having the eigenvalues
$\lambda_1=-\frac{J}{4}, \lambda_2=-\frac{J}{4}, \lambda_3=\frac{J}{4}, \lambda_4=\frac{J}{4}$\\
The corresponding eigenvectors are:
\begin{equation}
\vert \phi_1\rangle=\vert 00\rangle
\label{eqn:phi1}	
\end{equation}
\begin{equation}
\vert \phi_2\rangle=\vert 11\rangle
\label{eqn:phi2}	
\end{equation}
\begin{equation}
\vert \phi_3\rangle=-\vert 01\rangle+\vert 10\rangle
\label{eqn:phi3}	
\end{equation}
\begin{equation}
\vert \phi_4\rangle=\vert 01\rangle+\vert 10\rangle
\label{eqn:phi4}	
\end{equation}
The reduced density matrix is given by
$$\rho_{e}=\frac{1}{Z}\sum_{i}e^{-\beta \lambda_i}\vert \phi_i\rangle \langle \phi_i\vert,$$
where $Z$ is the partition function given by
\begin{equation}
Z=3e^{\frac{J}{4k_BT}}+e^{-\frac{3J}{4k_BT}}
\label{eqn:partitionfunction}	
\end{equation}
$\beta=1/k_BT$ and $k_B$ is the Boltzmann constant.\\
In matrix form

\begin{equation}\label{eqn:reddenmatrix}
\mathbf{\rho}_{e}=\frac{1}{Z}\left(\begin{array}{cccc}
v & 0 & 0 & 0\\
0 & w & z & 0\\
0 & z^* & w & 0\\
0 & 0 & 0 & v
\end{array}\right),
\end{equation}
with
$$v=\frac{1}{Z}e^{\frac{\beta J}{4}},$$
$$z=\frac{1}{Z}e^{-\frac{\beta J}{4}}\mathrm{sinh}\left(\frac{\beta J}{2} \right)$$

In terms of the correlations between the spins in an isotropic system \cite{wang}
$$v=\frac{1}{4}(1+4\langle s_{i}^{z}s_{i+1}^{z}\rangle),$$
$$z=\langle s_{i}^{x}s_{i+1}^{x} \rangle + \langle s_{i}^{y}s_{i+1}^{y} \rangle,$$

From equation \ref{eqn:distance}, it can be seen that the system will present entanglement when
$$v<\vert z \vert$$
In terms of the system's partition function:
\begin{equation}
v=\frac{1}{Z}e^{\frac{J}{4k_BT}},
\label{eqn:v}
\end{equation}
\begin{equation}
z=\frac{1}{Z}e^{-\frac{J}{4k_BT}}\mathrm{sinh}\left(\frac{J}{2k_BT} \right)
\label{eqn:z}
\end{equation}
From equations \ref{eqn:distance}, \ref{eqn:v} and \ref{eqn:z}, we get the distance between the states ($\mathcal{D}$) entanglement measure, $\mathcal{E}(\rho_e)$, for a dimer system as:
\begin{equation}
\mathcal{E}(\rho_e)=\mathrm{max}\left[0,\left(\frac{1-3e^{\frac{J}{k_B T}}}{1+3e^{\frac{J}{k_B T}}} \right) \right]
\label{eqn:distancefinal}
\end{equation}
From equation \ref{eqn:distancefinal}, it can be seen that the entanglement, $\mathcal{E}(\rho_e)$, is zero for ferromagnetic dimers ($J>0$). Furthermore, the entanglement temperature, T$_E$, at which entanglement goes to zero is:
\begin{equation}
T_E=-\frac{J}{k_B\mathrm{ln}3}
\label{eqn:enttemp}
\end{equation}

\subsubsection{Equivalence between distance between the states entanglement and Concurrence}
Wootters et al. \cite{wootters} calculated the concurrence of a spin dimer system as:
\begin{equation}
C=\mathrm{max}\{0, \lambda_1-\lambda_2-\lambda_3-\lambda_4\}
\label{eqn:concurrence}
\end{equation}
where $\lambda_i$ are the square roots of the eigenvalues of $\rho \tilde{\rho}$ in descending order such that $\tilde{\rho}$ is the spin-reversed density matrix, defined by $\tilde{\rho}=(\sigma^y \otimes \sigma^y)\rho^{T}(\sigma^y \otimes \sigma^y),$ $(\cdots)$ and the supercript $T$ indicates transposition.\\
Thus, the concurrence was found to be
$$C=2\mathrm{max}(0,\vert z \vert-v)$$
which is exactly the same as equation \ref{eqn:distance}.

\subsubsection{$\mathcal{E}(\rho_e)$ in terms of magnetic susceptibility}
To find realistic systems where bipartite entanglement exists at finite temperatures, it would be useful if the distance between the states measure, $\mathcal{E}(\rho_e)$, could be expressed in terms of experimentally accessible quantities like magnetic susceptibility and specific heat. In order to relate bipartite entanglement in a dimer with magnetic susceptibility, we use the Bleaney-Bowers equation given by \cite{bleaney}:
\begin{equation}
\chi= \frac{2N_A(g\mu_B)^2}{k_BT\left(3+e^{\frac{-J}{k_{B T}}}\right)}
\end{equation}
where $N_A$ is the Avogardo number, $g$ is the L\'{a}nde's factor and $\mu_B$ is the Bohr magneton. The concurrence for such a system was calculated by Aldoshin et al. \cite{aldoshin} as:
\begin{equation}
C=\mathrm{max}\left[0, 1-\frac{3\chi}{\chi_{Curie}}\right]
\label{eqn:concurrencedimer}
\end{equation}
where
$$\chi_{Curie}=\frac{N_A(g\mu_B)^2}{k_{B}T}$$ 
From the above section, we found the equality of the distance between the state measure, $\mathcal{E}(\rho_e)$, and concurrence $C$. Hence, equation \ref{eqn:concurrencedimer} becomes:
\begin{equation}
\mathcal{E}(\chi)=\mathrm{max}\left[0, 1-\frac{3\chi}{\chi_{Curie}}\right]
\label{eqn:distsusdimer}
\end{equation} 
\subsubsection{$\mathcal{E}(\rho_e)$ in terms of specific heat}
The specific heat per mole of a system in a canonical ensemble with partition function $Z$ can be written as: 
\begin{equation}\label{eqn:partition}
    c_m = \frac{\partial( {k_BNT^2\frac{1}{Z}\frac{\partial Z}{\partial T}})}{\partial T}
\end{equation}
For a magnetic spin 1$/$2 dimer system having a partition function $Z$ given by equation \ref{eqn:partitionfunction}, the magnetic specific heat becomes: 
\begin{equation}\label{eqn:Cm}
    c_{Mag}(T) =12k_BN\left(\frac{\beta J}{2}\right)^2\frac{e^{\beta J}}{(1+3e^{\beta J})^2}
\end{equation}
From equation \ref{eqn:distancefinal}, $e^{\beta J}$ can be expressed in terms of distance between the states as:
\begin{equation}\label{eqn:betaJ}
    e^{\beta J} = \frac{1-\mathcal{E}}{3(1+\mathcal{E})}
\end{equation}
Substituting equation \ref{eqn:betaJ} in equation \ref{eqn:Cm}, the distance between the state measure, $\mathcal{E}(\rho_e)$, can be expressed in terms of specific heat as:
\begin{equation}\label{eqn:entcv}
     \mathcal{E}(C_{Mag})= 
\begin{cases}
    \mathrm{max}\left[0,\sqrt{1-\frac{4C_m}{k_BN(\beta J)^2}} \right],& \text{for } T < T_E\\
        0,              &  \text{for }    T\ge T_E
\end{cases}
\end{equation}

\section{Computational details}
Interatomic distances and monomeric form of the crystal geometry were calculated with density functional theory (DFT)\cite{g16} using Gaussian 16 program package without any constriant on the geometry \cite{g16}. The structural parameters (atomic coordinates, inter-atomic distances etc.) obtained from SCXRD measurements were input to the Gaussian 16 program. DFT calculations have been done on the crystal for geometric optimization and simulation of Raman and IR spectrum by Becke three-parameter Lee Yang Parr functional correlation (B3LYP), a well-known functional in DFT for calculating exchange-correlation energy \cite{becke,lyp}. To correctly incorporate the effect of a metal centre (Cu$^{2+}$) and organic ligands (oxalate and acetate), we used a dual basis, with LANL2DZ for the metal centre and 6-311++G(d,p) for the rest of the atoms (organic part) \cite{abkari,fernando,georgeCRAT,sharath}. Energy minimized monomer unit obtained from DFT calculations was used to compute harmonic infrared vibrational frequencies as well as Raman frequencies and for the structural analysis. The calculations converged to an optimized geometry since there were only real harmonic vibrational wavenumbers, revealing the localization of energy minima. To study the chemical bonding and the magnetic interaction pathway, delocalisation indexes were calculated. Delocalisation index measures the extent of electron's delocalisation \cite{bertolotti} from one atomic space to another atomic space due to interatomic interactions. Optimised geometry from DFT calculation has been used to calculate electron density by Quantum Theory of Atoms In Molecule (QTAIM) using Multiwfn 3.8 software\cite{multiwfn}.
\section{Experimental section}
\subsection{Materials and General Methods}
The materials required for the synthesis of copper acetate were purchased commercially and used as-received: 4-Aminopyridine (Merck-Aldrich, 99$\%$), CuCl$_2$ (Alfa Aesar, 97$\%$), Acetic acid (Merck-Aldrich, reagent grade). The crystals were grown in Memmert's incubator (Model No. IPP30 PLUS 32L).  Thermogravimetric analysis (TGA) and differential scanning calorimetric analysis (DSC) measurements were done in TA Instruments thermal analyzer (Model no. SDT Q600) till 650 $^{\circ}$C at a heating rate of 5$^{\circ}$C/min under nitrogen environment. Fourier-transform infrared spectroscopy (FTIR) in the range of 400 cm$^{-1}$ - 4000 cm$^{-1}$ with 0.5 cm$^{-1}$ resolution, was done on Shimadzu’s spectrometer (Model No. IRPrestige-21) using the KBr pellet technique. Raman spectra were recorded with T64000 spectrophotometer using the 514.5 nm Ar–Kr laser excitation. Magnetisation measurements were done on a 3.4 mg crystal on Quantum Design's physical property measurement system (PPMS)-Model Evercool-II, in the temperature range 1.8 K to 400 K. The measurements were done in the zero field cooled (ZFC) state where the sample was cooled to the lowest temperature in zero field and then a field applied. Magnetisation measurements were performed while warming up the sample. Specific heat measurements were also done using PPMS in the temperature range from 2 K to 270 K. For the measurements, the platform was thermally attached to a temperature bath, and the sample, which has a mass of 3 mg, was thermally connected to it using Apiezon N grease. Specific heat was measured using the thermal relaxation approach, which involves applying a short heat pulse to the sample and extracting specific heat from exponential curve fits to the heating and cooling data.
\subsection{Synthesis}

\begin{figure}
\begin{center}
\includegraphics[width=1\textwidth]{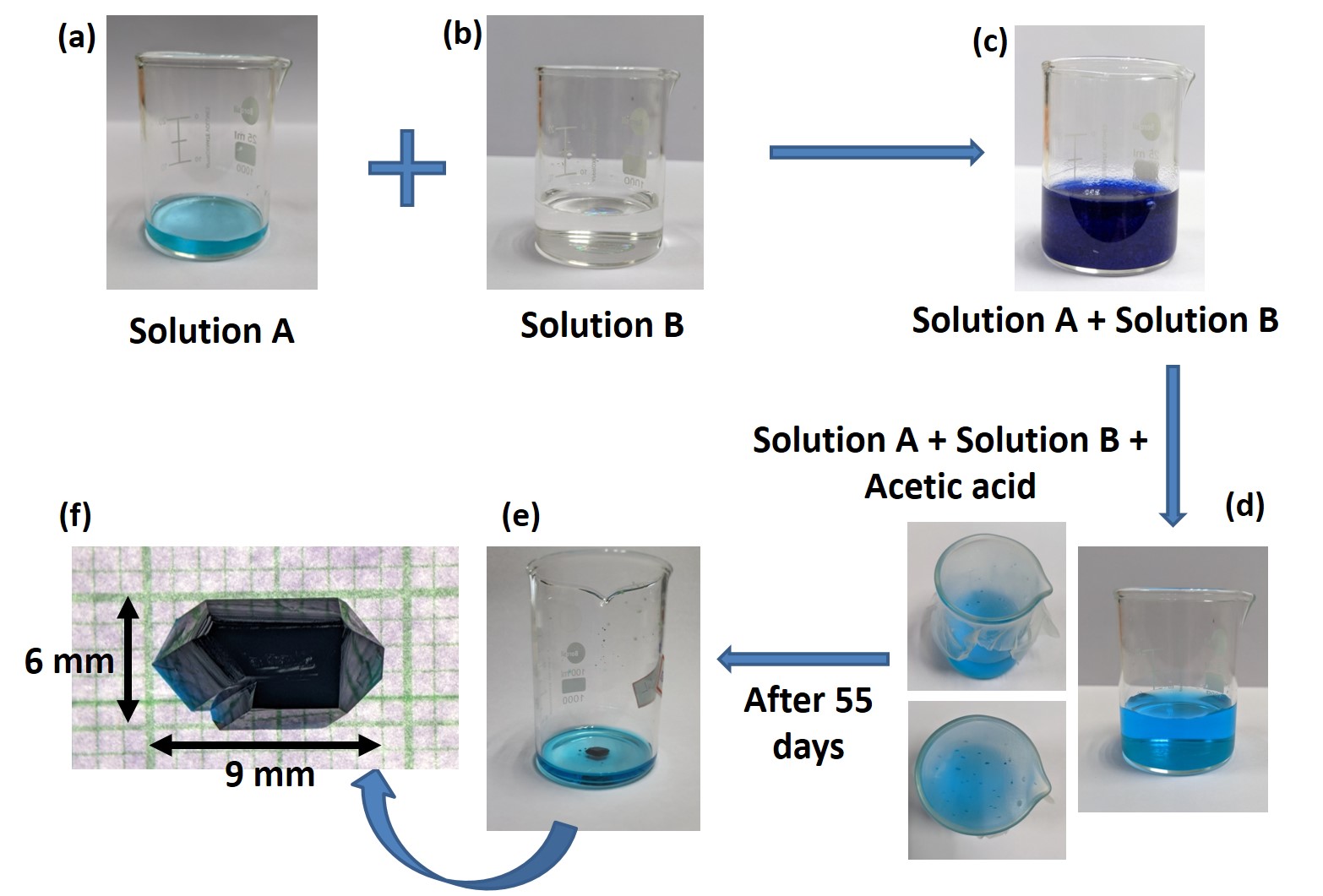}
\caption{(colour online) (a) Solution A of CuCl$_2$ in water (b) Solution B of 4-Aminopyridine in water. (c) Mixture of solutions A and B. (d) Growth solution after adding acetic acid to the mixed solutions A and B. (e) Crystal formation in growth solution. (e) A large sized crystal with hexagonal morphology shown on a graph paper for comparison.}
\label{fig:synthesis}
\end{center}
\end{figure}

Single crystals of copper acetate were synthesised using the technique of slow evaporation. The technique entails the slow evaporation of solvent from a solution until supersaturation resulting in crystallisation of excess solute \cite{zaitseva}. In our technique of slow evaporation, the solvent had two components: (i) a solution A of CuCl$_2$ dissolved in distilled water and (ii) a solution B of 4-Aminopyridine dissolved in distilled water. It was found that as soon as the two solutions A and B were mixed together, the solutes in each solution precipitated resulting in no crystal formation (see Fig. \ref{fig:synthesis} (c)). To inhibit the precipitation, we decided to add an acidic medium to the solution that would give H$^+$ ions to the system. Since acetic acid is a weak acid with a pK$_a$ of 4.76, we added acetic acid to the solution. As expected, after addition of acetic acid, the precipitation of the solvents was arrested and a clear blue solution formed (Fig. \ref{fig:synthesis} (d)). This solution was, then, left for slow evaporation.\\
To initiate slow evaporation, solution A was made by dissolving 0.17g of CuCl$_2$ in distilled water. Similarly, solution B was made by dissoving 0.37g of 4-Aminopyridine in distilled water. The two solutions were, then, mixed together in a beaker. Small volumes of acetic acid was added slowly to the mixed solution A and B while constantly stirring until the precipitates dissolved completely and a clear blue solution obtained as shown in Fig. \ref{fig:synthesis} (d). The obtained blue solution was allowed to evaporate slowly at room temperature in an incubator with the top covered with paraffin and few holes puntured on the paraffin. After 40 days, the first tiny crystal nucleated. With the passage of time, this crystal increased in size due to addition of secondary crystallites. After another 15 days, the crystal growth process ended with the formation of a large single crystal (see Fig. \ref{fig:synthesis} (e)) implying that the initial supersaturation attained in the mixture of solutions A and B was low \cite{zaitseva}. The crystal growth run gave us 63$\%$ yield. The morphology of the large sized crystals was hexagonal (see Fig. \ref{fig:synthesis} (f)). The size of each side of the hexagon was $\sim$ 3 mm with a thickness of $\sim$ 0.3 mm, while the dimension of the rectangular crystals was 10 mm x 10 mm x 0.3 mm. The obtained crystals were washed in minimal amount of cold distilled water and the remaining excess solvent blotted on a clean tissue paper. The large single crystal was cut to an approximately rectangular shape of dimension 1 cm x 1 cm x 0.3 cm for other measurements. Our objective in taking 4-Aminopyridine as the starting reagent for the crystal synthesis was to use it either as a bridging ligand or a side ligand in order to engineer an exchange interaction between the Cu$^{2+}$ ions. However, structure solution (discussed below) revealed that copper acetate did not contain any aminopyridine.  
\subsection{X-ray data collection and structure determination} 
A tiny prismatic single crystal of dimensions 0.25 x 0.12 x 0.05 mm$^3$ was mounted on the goniometer of Bruker's Kappa APEX II CCD diffractometer equipped with graphite-monochromatized Mo-K$_\alpha$ radiation having a wavelength $\lambda$ = 0.71073 \AA~ at room temperature (296(2) K). The intensity data was collected using $\omega$ and $\phi$ scans with frame width of 0.5$^{\circ}$. The frame integration and data reduction were performed using Bruker's SAINT/XPREP software \cite{bruker}. Multi-scan absorption corrections were applied to the data using SADABS (Bruker 1999) \cite{sadabs} program. Intensity distribution indicated a monoclinic structure with the space group as C 2$/$c, which was confirmed by a successful refinement. Accurate unit cell parameters and orientation matrix were determined by least-squares treatment of the setting angles of 5689 reflections, of which 1233 reflections were independent, in the 2.947$^{\circ} \le 2\theta \le$ 24.996$^{\circ}$ range. The minimum and maximum normalized transmission factors were 0.772 and 0.926. Atomic positions were located by Direct Methods with the structure solution program SHELXT \cite{shelxt} and were then refined by full-matrix least-squares calculations based on F2 using the program SHELXL \cite{shelxl}. Selected crystallographic data is given in the Table \ref{table:xrd}. From the Table \ref{table:xrd}, it can be seen that the goodness of fit R has a low value of 1.117 indicating an extremely good quality of the grown crystal.

\begin{table}
	\begin{center}
		\caption{Selected crystallographic data of C$_8$H$_{16}$Cu$_2$O$_{10}$ obtained from single crystal X-ray diffraction.}
		\begin{tabular}{|cc|}
			\hline 
			\bf {Empirical formula}&\bf{C$_8$H$_{16}$Cu$_2$O$_{10}$} \\ 
			\hline 
			Formula weight&399.31\\ 
			
			Crystal System&Monoclinic\\ 
			
			Space Group&C 2/c\\ 
			
			Unit cell dimensions& a = 13.1585(8) \AA \\
			& b = 8.5575(5) \AA \\
			&c = 13.8517(9) \AA \\
			&$\alpha$ = 90$^\circ$\\ 
			
			&$\beta$ = 117.035(2)$^\circ$ \\ 
			
			&$\gamma$ = 90$^\circ$  \\  
			
			Volume&1389.32(15) \AA$^{3}$  \\ 
			
			Z&4  \\ 
			
			Density (calculated)&1.909 Mg/m$^3$ \\ 
			
			Absorption coefficient&3.106 mm$^{-1}$ \\
			
			Index ranges&-15$<$=h$<$=15, -10$<$=k$<$=10,  \\
			& -16$<$=l$<$=13\\
			
			Goodness-of-fit on F$^{2}$& 1.117 \\ 
			
			Final R indices [I $>$ 2$\sigma$(I)]&R1 = 0.0221, wR2 = 0.0550\\ 
			
			R indices (all data)&R1 = 0.0265, wR2 = 0.0571 \\ 
			
			Extinction coefficient&0.0187(7) \\
			
			Largest diff. peak and hole&0.421 and -0.227 e.$\AA ^{-3}$  \\
			\hline
		\end{tabular}  
		
		\label{table:xrd}
	\end{center} 
	
\end{table}

\section{Result and Discussion}
\subsection{Crystal structure} 
\begin{figure}
	\begin{center}
		\includegraphics[width=1\textwidth]{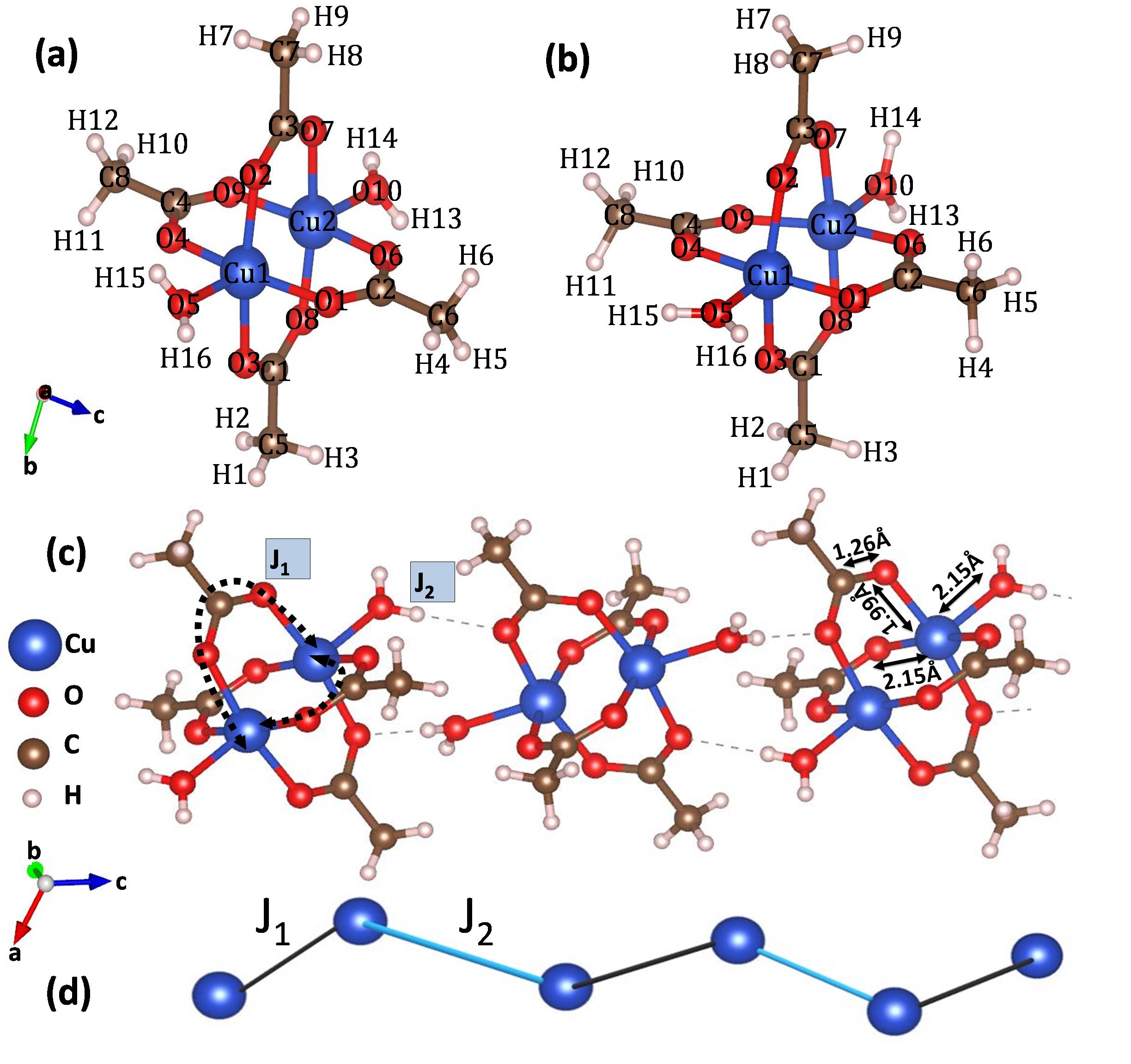}
		\caption{(colour online) (a) Asymmetric unit of C$_8$H$_{16}$Cu$_2$O$_{10}$ obtained from single crystal X-ray diffraction. (b) Optimized geometry of C$_8$H$_{16}$Cu$_2$O$_{10}$ generated by the dual basis (see text for details). (c) Atomic labeling is shown.}
		\label{fig:asymmetric}
	\end{center}
\end{figure}

Copper acetate crystallises in the monoclinic system with half the number of molecules in the asymmetric unit. The two halves are related to each other by a centre of inversion. Fig. \ref{fig:asymmetric} (a) shows the asymmetric unit obtained from SCXRD measurements. From the figure, it can be seen that the asymmetric unit contains 2 Cu atoms, Cu1 and Cu2. Each Cu atom is penta-coordinated with oxygen atoms: Cu1 with O1, O2, O3, O4 and O5; Cu2 with O6, O7 O8, O9 and O10. Four of these O atoms arise from the acetate ligand CH$_3$COO$^-$. It is interesting to note that out of the two O atoms in a given acetate ligand, one (for instance, O4) is co-ordinated with Cu1 atom while the other O atom (O9) is co-ordinated with Cu2 atom. The fifth penta-coordinated O atom with Cu atom arises from a water molecule: H15-O5-H16 with Cu1 and H13-O10-H14 with Cu2 atom.
\begin{table}
	\centering
	\begin{tabular}{||c c c c c c|| } 
		\hline
		& Experiment & Theory &    &Experiment & Theory\\
		\hline 
		Bond length[\AA] & & &Bond angle[$^{\circ}$] & &\\
		\hline
		Cu1-Cu2 &  2.61(7) &  2.73 & O2-Cu1-O4	&  89.21(7)& 89.14\\
		Cu1-O1	&  1.95(19) &  2.04 & O3-Cu1-O5	&  93.05(8) & 93.41\\
		Cu1-O2	&  1.98(15)&  1.99 & O3-Cu1-O1	&  89.21(7)& 89.14\\
		Cu1-O3	&  1.99(15) & 1.99 & O1-Cu1-O2       & 90.98(7)& 89.72\\
		Cu1-O4	&  1.93(3) &  2.04 & O4-Cu1-O5       & 97.68(9)& 99.52\\
		Cu1-O5	&  2.15(3) &  2.14 & O2-Cu1-O5	& 98.29(8)& 93.41\\
		Cu2-O6	&  1.93(3) &  1.99 & O2-Cu1-O3	& 168.63(10)& 173.17\\
		Cu2-O7	&  1.99(15) &  2.04 &O1-Cu1-O4	& 168.66(10)& 160.95\\
		Cu2-O8	&  1.98(15) &  2.04 & Cu1-O5-H15	& 134.00(3)& 123.10\\
		Cu2-O9	&  1.95(19) & 1.99 & O7-Cu2-O8	& 168.63(10)&  160.93\\
		Cu2-O10	&  2.15(3) & 2.14 & O9-Cu2-O6	& 168.66(10)& 173.16\\
                &          &      & O7-Cu2-O10	& 93.05(8)& 99.53\\
                &          &      & O8-Cu2-O10	& 98.29(8)& 99.53\\
                &          &      & O7-Cu2-O6       & 90.25(8)& 89.71  \\
                &          &      & O7-Cu2-O9       & 90.98 (8)& 89.15  \\
                &          &      & O8-Cu2-O6       & 97.34(8)& 89.15 \\
                &          &      & O9-Cu2-O8       & 90.98(7)& 89.71 \\
                &          &      & Cu2-O10-H13	& 134.00(3) & 123.10 \\
                &          &      & Cu2-O10-H14     & 114.00(3) & 123.10 \\

		\hline
	\end{tabular}
	\caption{Comparison of experimentally obtained SCXRD crystallographic data of C$_8$H$_{16}$Cu$_2$O$_{10}$ with that obtained from DFT calculations.}
	\label{table:SCXRDDFT}
\end{table}
Four acetate ligands CH$_3$COO$^{-}$, each with a net negative charge warrants that the oxidation state of Cu in copper acetate is Cu$^{2+}$, ensuring that the total charge on an asymmetric unit is zero making the molecule neutral. Lattice parameters obtained from SCXRD data are given in the Table \ref{table:SCXRDDFT}.\\
Monomeric unit pertaining to the optimised geometry of copper acetate obtained from the DFT calculations employing the dual basis set as described in the above section is shown in Fig. \ref{fig:asymmetric} (b). From Figs. \ref{fig:asymmetric} (a) and (b), it is apparent that the experimentally obtained asymmetric unit and the monomeric unit obtained from DFT calculations are equivalent. Bond lengths corresponding to the magnetic copper atom are given in the Table \ref{table:SCXRDDFT}. From the table, it can be seen that an excellent match of the experimentally obtained bond lengths exists with those of the theoretically obtained one resulting in negligible discrepancies with the experimental geometry in the bond lengths. It is to be remembered that the calculations were performed without any constraint on the geometry unlike \cite{bertolotti} where calculations were done on a fixed geometry to minimise the differences in the computationally and experimentally obtained bond lengths. This suggests that our dual basis incorporating both the metal and organic parts is a good basis to describe the structure of copper acetate. Bond angles of copper with oxygen of the kind O-Cu-O is also very good and those with oxygen and hydrogen atoms of the kind Cu-O-H is reasonably good. Considering the fact that the DFT calculations are done on a single molecule of copper acetate that is assumed to be in a gaseous state, the good match with experimentally and theoretically obtained data implies that the actual crystalline state of copper acetate may comprise one dimensional chains of copper that are magnetically isolated in the other two directions due to weak interchain interactions.\\ 
The molecular chain structure of copper acetate obtained from the SCXRD data is shown in Fig. \ref{fig:asymmetric} (c). The crystal is composed of a chain of units of copper linked by H-bonds along the c-axis. Each unit in copper acetate is bound to the neighbouring unit through two hydrogen bonds O–H···O,
involving acetate oxygen atoms and water molecules. A unit of copper comprises two copper atoms that are bonded through four acetate groups to form a dimer. The bonds formed by the four acetate groups give the well-known paddlewheel structure to the molecule \cite{li}. Each of the central copper atom is connected to four oxygen atoms from the acetate group and one oxygen atom from the water molecule. The H-bond is shared with the water molecule and adjacent acetate group in the next molecule (see Fig. \ref{fig:asymmetric} (c)). Cu ions in the molecule are separated by 2.61 \AA,  and the intermolecular distance along the c-axis is 5.2 \AA.  Dimer atoms are connected by H-bond along the b-axis with a separation of 7.2 \AA~ which is larger than the separation in the c-axis. Since the Cu-Cu distance in copper acetate (2.619 \AA) is very close to the Cu-Cu distance in metallic copper (2.56 \AA), a direct exchange between the two copper atoms seems a big possibility. However, the Cu-O(CO) is distance (2.616 \AA) is smaller than the sum of the ionic radii of copper and oxygen atoms implying a covalent bond in copper acetate and a possible superexchange between the copper atoms consequently \cite{kuzmina}. In order to confirm if the primary source of magnetic exchange between the copper atoms is via direct exchange or is mediated via oxygen atoms through superexchange, we calculated the delocalisation index, $\delta(A,B) = 4\int_{\Omega(A)}dr_{1}\int_{\Omega(B)}dr_{2}\rho(r_{1},r_{2})-2N(A)N(B)$, where $\rho(r_{1},r_{2})$  is the two-particle density for electrons of parallel spin \cite{fradera}. $\delta(A,B)$ denotes the extent of delocalisation of electrons in an atom A towards the electrons of another atom B \cite{koritsanszky} such that a value of unity is attained for bond formation.

\begin{table}[h!]
\centering
 \begin{tabular}{||c c|| } 
 \hline
Bond pairs & B3LYP \\
 	\hline 
      &  6-311++G(d,p)-LANL2DZ\\
 \hline
Cu-Cu    & 0.527 \\
Cu-Oac	 &  0.809 \\
Cu-Ow   &  0.630 \\
O-Hw	 &  0.865 \\
C-C   	&  1.043 \\
C-Oac	&  1.600 \\
C-H 	&  0.882 \\

 \hline
 \end{tabular}
 \caption{Delocalization index for copper acetate calculated using DFT B3LYP with dual basis set 6-311++G(d,p)-LANL2DZ.}
 \label{table:DI}
\end{table}

Table \ref{table:DI} depicts the details of the calculated  $\delta(A,B)$ for interatomic interactions present in the copper acetate in the basis set, B3LYP/6-311++G(d,p)-LANL2DZ. The estimated DI for C-Oac ($\delta(C,O)$ = 1.601) is comparable to the typical DI for $\pi$ and $\sigma$ bonds in the acetate molecule (O-C=O), which is 1.5. Similar to what is expected for a covalent bond, the DI values of the O-H, C-C, and C-H bonds are close to 1 \cite{Tian}. For the C-O bond in the acetate group, there is also a covalent bond with $\delta(C,O)$ = 0.809, which is again close to 1. With $\delta(Cu,O)$ = 0.809, the covalent bond between copper and oxygen is also close to 1. Given that the interactions in copper acetate are of metal-ligand kind, DI values are lower than that of strong covalent bonds in C-C or C-O. The important point is that the $\delta(Cu,Cu)$ for direct bond yields even smaller values equal to 0.527, which is comparable with the values of similar molecules when no bond pathway is observed \cite{Holladay}. This low value of the $\delta(Cu,Cu)$ bond in comparison to the other  $\delta(Cu,O)$, $\delta(C,C)$, $\delta(C,O)$ and $\delta(C,H)$ bonds shows that direct electron sharing between two Cu atoms is unlikely. This also eliminates the possibility of direct magnetic interaction between Cu atoms, providing only the super exchange mediated by Cu-O bonds as the only path to dimer formation.\\
Since the value of Cu1-O3 and Cu2-O7 is the same at 1.99 \AA ~ and Cu1-O5 and Cu2-O10 is same at 2.15 \AA ~ (refer Table \ref{table:SCXRDDFT}), there are two possible exchange pathways for the intra-dimer connection, namely, Cu1-O3-H14 (at an angle of 116.9$^{\circ}$) and Cu2-O7-H15 (at an angle of 116.9$^{\circ}$). It is clear from the polyhedral chain structure that the dimers (Cu2-Cu1 and Cu2-Cu1) are not in the same plane along the c-axis and are making a dihedral angle of 94.42(4)$^{\circ}$. These nonplanar bonds give a zig-zag chain structure to the crystal along the c-axis. Due to the two different interactions in intra-dimer and inter-dimer copper ions, a dimeric copper chain is formed, as shown in Fig. \ref{fig:asymmetric} (d) having very different values of the exchange coupling constants $J$ and $\alpha$$J$ such that $\alpha$$J$ is much smaller than $J$.

\begin{figure}
	\centering
	\includegraphics[width=1\textwidth]{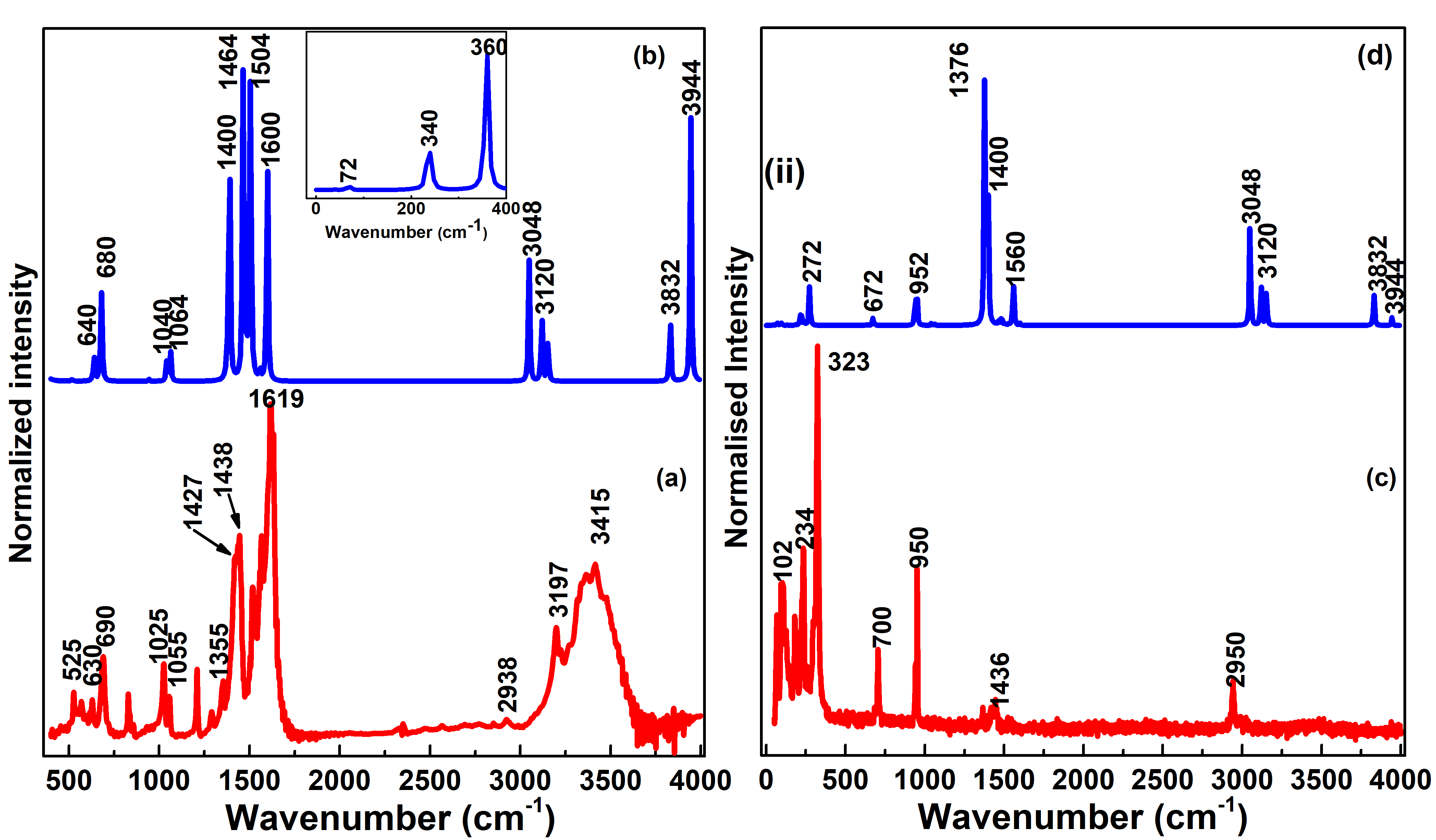}
	\caption{(a) Red curve is the experimentally obtained room temperature (298 K) FTIR spectrum of C$_8$H$_{16}$Cu$_2$O$_{10}$. (b) Blue curve is the theoretically simulated spectrum using the dual basis. (c) Raman spectrum of C$_8$H$_{16}$Cu$_2$O$_{10}$ obtained  experimentally at room temperature (red curve) and (d) Theoretically simulated (blue curve). Frequencies corresponding to relevant vibrations are marked.}
	\label{fig:FTIRRaman}
\end{figure}

\subsubsection{FTIR and Raman spectroscopy}
In order to understand how the theoretically obtained infra red and Raman frequencies compare with the experimentally obtained one, we measured FTIR and Raman spectra of copper acetate and plotted them in Fig. \ref{fig:FTIRRaman}. Red curves in Fig. \ref{fig:FTIRRaman} depict the experimentally obtained spectra while the blue correspond to the theoretically simulated spectra  computed on a monomeric unit using the mixed basis (see section above). From the monomeric unit shown above (see Fig. \ref{fig:asymmetric} (a-(c))), acetate, water and Cu-O bonds are expected to contribute primarily to the FTIR and Raman frequencies. So, the peaks corresponding to the acetate ligand are observed at frequencies of $\nu_1$ = 2938 cm$^{-1}$ (sym. CH$_3$ stretch); $\nu_2$ = 1427 cm$^{-1}$ (sym. C-O stretch); $\nu_3$ = 1355 cm$^{-1}$ (sym. CH$_3$ bend); $\nu_4$ = 1619 cm$^{-1}$ (asym. C-O stretch); $\nu_5$ = 1438 cm$^{-1}$ (asym. CH$_3$ bend); $\nu_{6}$ = 1055 cm$^{-1}$ (CH$_3$ asym. bend); $\nu_{7}$ = 1025 cm$^{-1}$ (CH$_3$ rocking); $\nu_{8}$ = 690 cm$^{-1}$ (sym. COO bending) and $\nu_{9}$ = 630 cm$^{-1}$ (COO rocking) in the FTIR spectrum of Fig. \ref{fig:FTIRRaman} (a).\\
The FTIR spectrum of copper acetate also exhibits strong absorption in the regions of the stretching vibrations of coordinated water molecules as evident in the intense peaks at 3415 cm$^{-1}$ \cite{kuzmina}, 3197 cm$^{-1}$ and 1619 cm$^{-1}$ (scissoring (H$_2$O)) \cite{Kazuo} of Fig. \ref{fig:FTIRRaman} (a). The expected magnetic interaction between the two copper atoms mediated by the paddlewheel arrangement of Cu-O atoms (dotted curve in Fig. \ref{fig:asymmetric} (c)) should result in a peak in the FTIR spectrum at a Cu-O frequency. Indeed, the low intensity peaks at 525 cm$^{-1}$, 590 cm$^{-1}$ and 628 cm$^{-1}$ are observed in Fig. \ref{fig:FTIRRaman} (a) confirming the presence of Cu-O vibrations in copper acetate \cite{elango,narang}.\\
The theoretically simulated FTIR spectrum using the mixed basis shows a very good match of the calculated frequencies with the experimentally obtained ones as shown in Fig. \ref{fig:FTIRRaman} (b), proving the usefulness of the mixed basis for DFT calculations of metal-organic compounds. It is to be noted that some of the IR frequencies corresponding to Cu-O vibrations are present below 500 cm$^{-1}$ \cite{mathey}. Due to experimental constraints of our instrument that has a poor resolution of the frequencies below 500 cm$^{-1}$, we were unable to measure these frequencies. However, an indication of the contribution of these frequencies to the IR spectrum of copper acetate was obtained from the DFT simulated IR spectrum as shown in the inset of Fig. \ref{fig:FTIRRaman} (b) where frequencies at $\nu_s$ = 72 cm$^{-1}$ (Cu-Cu-O bending), $\nu_s$ = 340 cm$^{-1}$ (O-C-uO bending) and $\nu_s$ = 360 cm$^{-1}$ (sym. CuO stretch) are observed \cite{mathey}.\\
In order to confirm the low frequency bands associated with Cu-O, we performed Raman spectroscopy on copper acetate in the range of 50 cm$^{-1}$ to 4000 cm$^{-1}$ as shown in Fig. \ref{fig:FTIRRaman} (c). Vibrations of the acetate ligand are observed at $\nu_1$ = 1436 cm$^{-1}$ (asym. CH$_3$ bending); $\nu_2$ = 950 cm$^{-1}$ (sym. CCH$_3$ streching) and $\nu_3$ = 700 cm$^{-1}$ (sym. COO bending) \cite{mathey}. Unlike the FTIR spectrum, the Raman spectrum presents clear peaks in the low frequency range where frequencies corresponding to Cu-O vibrations are seen at $\nu_4$ = 323 cm$^{-1}$ (sym. CuO streching); $\nu_5$ = 234 cm$^{-1}$ (O-Cu-O bending) and $\nu_6$ = 102 cm$^{-1}$ (Cu-Cu-O$_w$ bending). Fig. \ref{fig:FTIRRaman} (d) shows the simulated Raman spectrum obtained using the mixed basis. As can be observed from the spectrum, a good match is obtained between the experimentally obtained and theoretically simulated spectra.\\ 

\subsubsection{Thermal Analysis}

\begin{figure}
	\centering
	\includegraphics[width=1\textwidth]{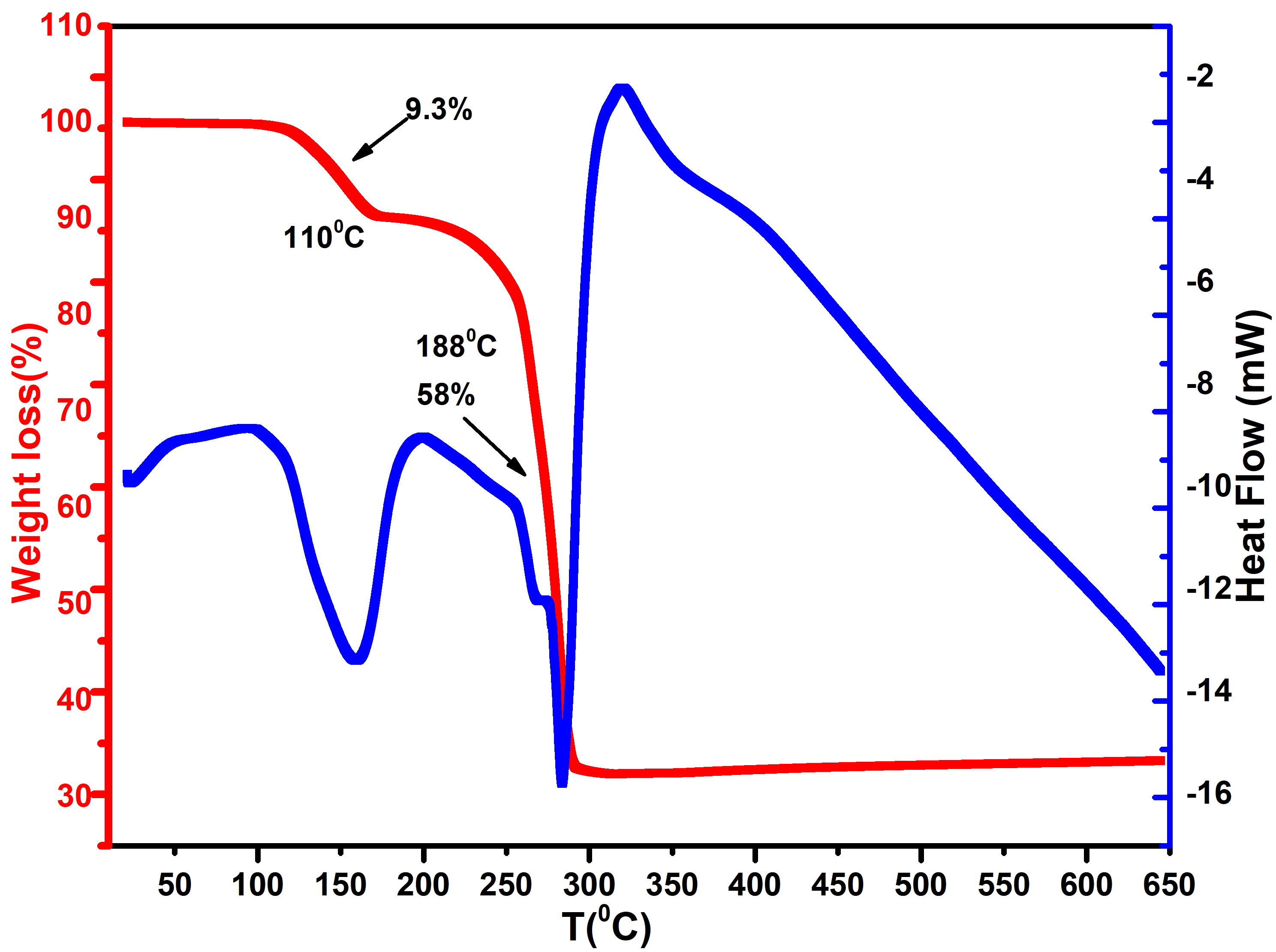}
	\caption{TGA (red) and DSC (blue) curves of copper acetate showing two distinct weight losses.}
	\label{fig:TGA}
\end{figure}

The copper acetate crystal was subjected to TGA and DSC measurements to ascertain it's thermal stability. The TGA data collected in the 20$^{\circ}$C to 640$^{\circ}$C temperature range is represented as a red curve in Fig. \ref{fig:TGA}. Blue curve in Fig. \ref{fig:TGA} shows the DSC measurement that was carried out simultaneously to determine the nature of the underlying change (endothermic or exothermic). It was found that the weight reduction happens in two stages. First, a weight loss of 9.3$\%$ at 110$^{\circ}$C is observed accompanied by a small endothermic peak. The amount of water molecules in the crystal, as a proportion of its weight, is discovered to be about 9.02$\%$. So, it is obvious that the weight loss is a result of dehydration. The slight difference between the measured mass loss and the absolute dehydration value predicted by theory for C$_8$H$_{16}$Cu$_2$O$_{10}$ suggests that other mechanisms causing mass loss take place in this temperature range. One of such possible events is the oxidation of the sample's surface to form copper acetate peroxides \cite{Zhenkun}. Impure nitrogen containing a small quantity of oxygen can make copper acetate peroxides since they just need a small amount of oxygen to form. Due to their instability, these peroxides start to break down at low temperatures (below 168$^{\circ}$C), which causes further mass loss between 110 to 180$^{\circ}$ besides the dehydration. Second, at 188$^{\circ}$C,  there is a significant weight loss of about 58$\%$, accompanied by a sizable endothermic peak in the DSC data. This weight loss could be caused by the disintegration of acetate links connecting the copper atoms in the dimer, which has a weight percentage of $\sim$ 59$\%$ \cite{Itab} when the four clusters forming the paddle wheel structure are considered. The sharp endothermic peak shows that except for the liberation of crystalline water at 110$^{\circ}$C, this crystal is stable up to 188$^{\circ}$C without any phase transition \cite{georgeCRAT}.

\subsection{Bipartite entanglement}
\subsubsection{via magnetic susceptibility} 
\begin{figure}[h!]
	\centering
	\includegraphics[width=1\textwidth]{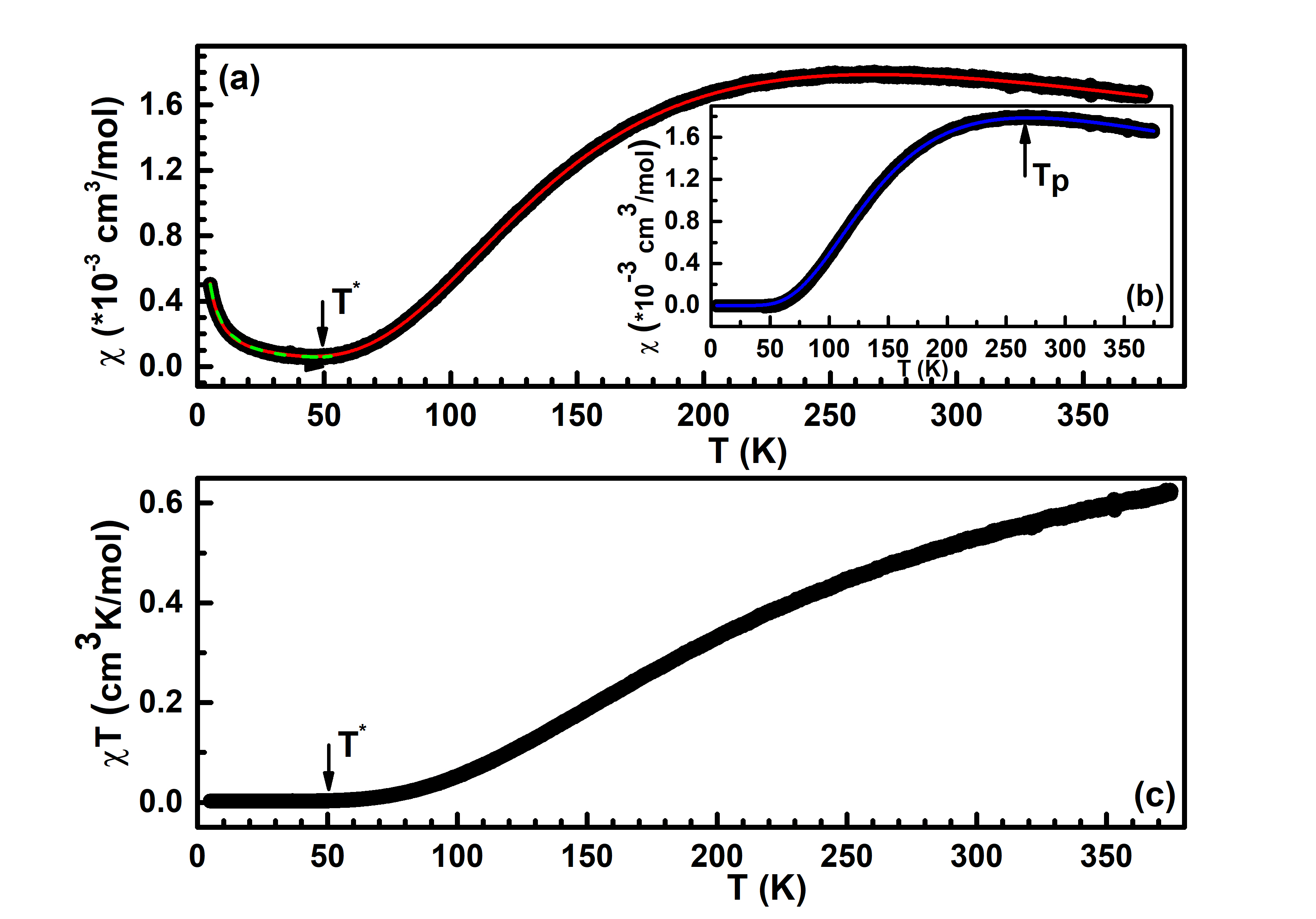}
	\caption{(a) Black filled circles represent the temperature variation of magnetic susceptibility at an applied field of 50 mT. Red curve is a fit to a 1D Heisenberg alternating chain model. Green dashed line is a fit to the equation $\chi$(T) = $\chi_0$ + $\chi_{CW}$(T) + $\chi_{spin}$(T). (b) Black filled circles represent the temperature variation of magnetic susceptibility after subtracting the uncoupled spin contribution to the data. Blue curve is a fit to the Bleaney Bowers expression (see text for details). The peak temperature T$_p$ is marked. (c) Temperature variation of $\chi$T.}
	\label{fig:susceptibility}
\end{figure}

From the delocalisation index calculations above as well as the IR and Raman measurements above, it is clear that the exchange interaction happens through the paddlewheel structure formed via the acetate group rather than the direct Cu-Cu exchange. In order to ascertain the formation of a dimer ground state and a consequent estimation of bipartite entanglement via the distance between the state measure, it is essential to measure magnetic susceptibility as well as specific heat of copper acetate. Black filled circles in Fig. \ref{fig:susceptibility} (a) denote the temperature variation of experimentally obtained magnetic susceptibility. From the graph, it is clear that a broad peak exists at a temperature $T_p \sim$ 260 K, similar to other published works \cite{mookerjee,gregson} and indicative of low dimensionality. The figure also presents a low temperature upturn that starts below $\sim$ 50 K and rises sharply below 20 K. Since the expected ground state is that of a dimer, such an upturn may arise due to uncoupled spins arising due to paramagnetic impurities \cite{He}. To confirm this, a fit to a one dimensional Heisenberg alternating chain model was done incorporating impurities due to uncoupled spins (para), alternation parameter ($\alpha$), Curie constant (CC) and exchange coupling constant (J/k$_B$) as fit parameters\cite{Landee}. Red curve in Fig. \ref{fig:susceptibility} (a) is the resultant fit with the obtained fit values as para = 0.27$\%$, $\alpha$ = 0.0017, CC = 0.95 and J/k$_B$ = 430 K. The vanishingly low value of $\alpha$ obtained as a result of an excellent fit to the data implies that the ground state of copper acetate is a dimer as expected. The percentage impurity spins in our crystal is quite low at 0.27$\%$, however, it is necessary to subtract this paramagnetic contribution in order to obtain the susceptibility of the dimer. To do this, we fitted the data below 50 K (green dashed line in Fig. \ref{fig:susceptibility} (a)) to $\chi$(T) = $\chi_0$ + $\chi_{CW}$(T) + $\chi_{spin}$(T) \cite{johnston,He} where $\chi_0$ is a temperature independent term, $\chi_{CW}$(T) = C/T-$\theta$ is the Curie-Weiss term arising due to magnetic impurity and $\chi_{spin}$(T) = aT$^{-1/2}$exp(-$\Delta$/k$_B$T) is the spin susceptibility of a one dimensional chain with a finite gap $\Delta$. The obtained fit parameters are as follows: $\chi_0$ = -9.4 x 10$^{-5}$ cm$^3$/mol, C = 0.023 cm$^3$K/mol, $\theta$ = -0.58 K and $\Delta$/k$_B$ = 352 K. The obtained $\chi_0$ value is close to the value obtained from closed shell diamagnetic susceptibility value -17.5 x 10$^{-5}$ cm$^3$/mol.

\begin{figure}[h!]
	\centering
	\includegraphics[width=1\textwidth]{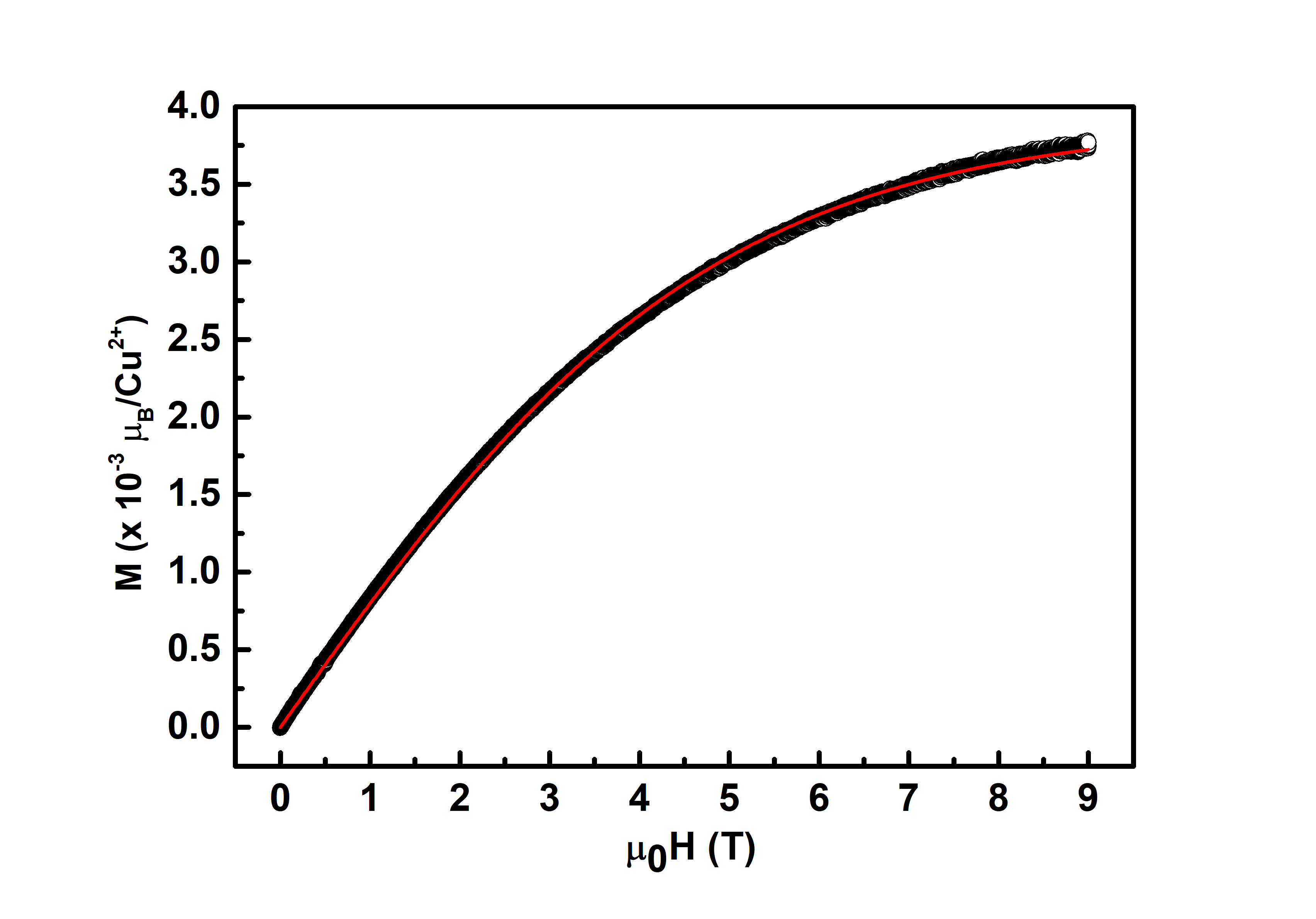}
	\caption{Black open circles represent the field variation of magnetisation while the red solid curve is a fit to the Brillouin function.}
	\label{fig:MH}
\end{figure} 

Black curve in Fig. \ref{fig:susceptibility} (b) is the magnetic susceptibility of copper acetate obtained after subtracting the impurity contribution arising due to paramagnetic impurities. As can be seen from the figure, the susceptibility vanishes to zero at low temperatures as expected for a dimer ground state due to formation of singlets. To confirm the dimer ground state, we fitted the susceptibility to Bleaney Bowers expression of the form $\chi$(T) = 2$Ng^2 \mu_B^2/k_B T$[3+exp($\Delta/k_BT$)] where symbols have their usual meaning. Blue dashed curve in Fig. \ref{fig:susceptibility} (b) is the fitted curve which is seen to overlap the experimentally obtained data, confirming the dimer ground state. The obtained fit parameters are $g$ = 2.17 and $\Delta/k_B$ = 426 K. A final confirmation of the presence of magnetic impurities in the system was obtained by doing a field dependent magnetisation measurement as shown in Fig. \ref{fig:MH}. Here, the Brillouin function\cite{Landee}:
\begin{equation}
    M_{mol}(B,T) = M_{sat}B_s (g\mu_BSB)
\end{equation}
\begin{equation}
    B_s(g\mu_BSB) = \frac{2S+1}{2S}coth(\frac{2S+1}{2S}\frac{g\mu_BSB}{k_BT})-\frac{1}{2S}coth(\frac{1}{2S}\frac{g\mu_BSB}{k_BT})
\end{equation}
shown in the red curve is found to fit the data perfectly well in the measured field range of 0 to 9 T. The obtained fit parameter values are M$_{sat}$ = 3.7 and g = 2.9 where M$_{sat}$ is the saturated magnetisation, S is spin and B is the applied magnetic field.

\begin{figure}[h!]
	\centering
	\includegraphics[width=1\textwidth]{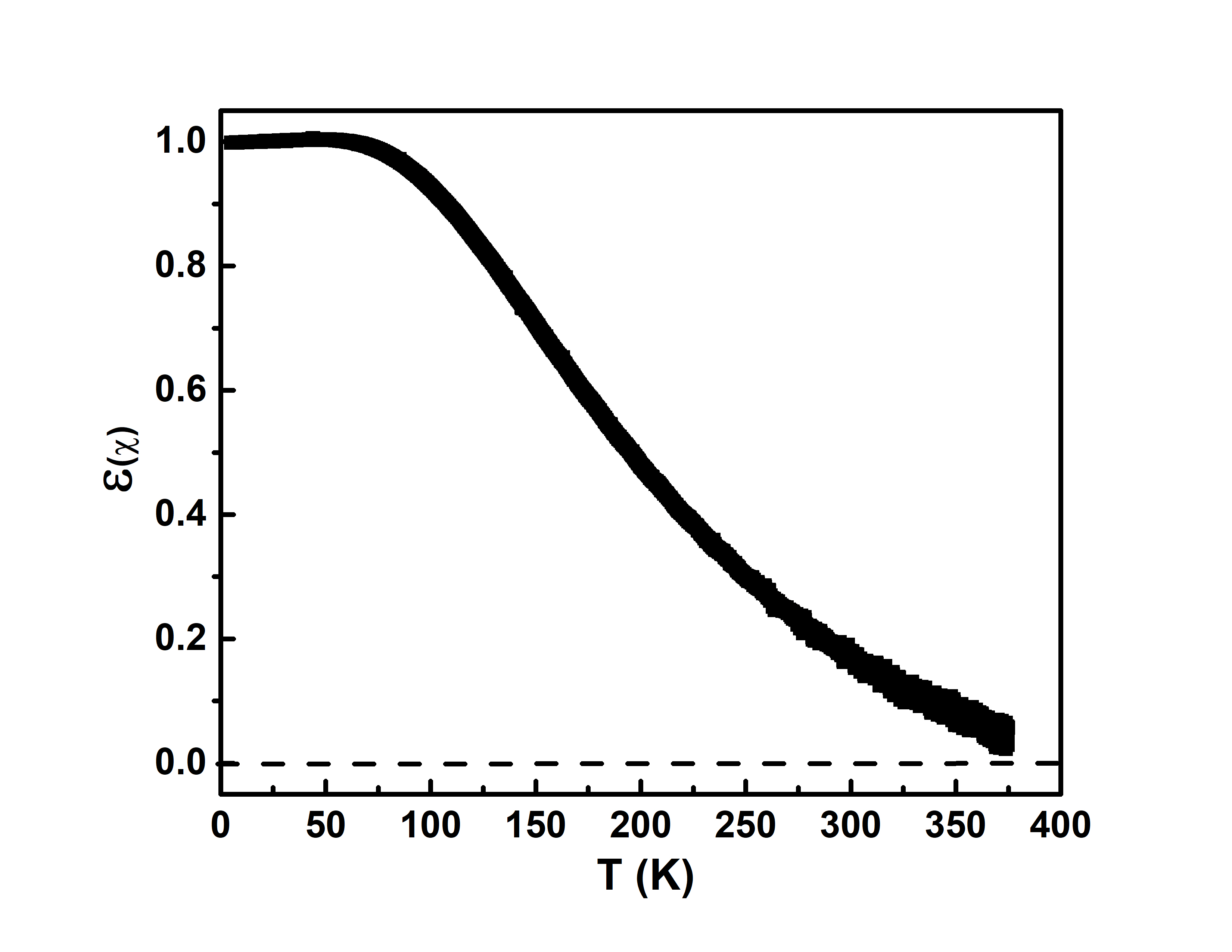}
	\caption{Temperature variation of distance between the states in terms of magnetic susceptibility, $\mathcal{E}$($\chi$).}
	\label{fig:entanglement}
\end{figure}

From equation \ref{eqn:concurrencedimer}, the distance between the states, $\mathcal{E}$, is obtained in terms of the product of susceptibility and temperature, so it is informative to plot the temerature variation of $\chi$(T)T as has been shown in Fig. \ref{fig:susceptibility} (c). The susceptibility values correspond to the ones from Fig. \ref{fig:susceptibility} (a) without any paramagnetic subtraction. As can be seen from Fig. \ref{fig:susceptibility} (c), $\chi$(T)T decreases montonically with temperature indicating antiferromagnetic correlations. Below $\sim$ 50 K, $\chi$(T)T is seen to become a constant, implying paramagnetic behaviour, consistent with the discussions above.\\
We now, plot the temperature variation of the distance between the states in terms of magnetic susceptibility, $\mathcal{E}$($\chi$), as shown by the filled black circles in Fig. \ref{fig:entanglement}. From the figure, it can be seen that $\mathcal{E}$($\chi$) starts from the maximum possible value of 1 \cite{conner} at the lowest measured temperature of 1.8 K and starts to decrease steadily with a further increase in temperature. The dimer given by Hamiltonian \ref{eqn:Hamiltoniandimer} has two energy levels +3J/2 and -J/2 separated by the energy gap $\Delta$, where the first level is a singlet and the second is a triply degenerate level. At absolute zero of temperature, the singlet state of copper acetate is completely populated making its spin entanglement correspond to that of a pure and maximally entangled state. As the temperature increases, the triplet state starts to get populated reducing the total entanglement of copper acetate. Very surprisingly, it retains a large finite value of 0.5 at a high temperature of 150 K. With a further increase in temperature, the entanglement value decreases further and reaches an extremely low value of $\sim$ 0.05 at 380 K. However, it is to be noted that the entanglement has still not reached a zero value. The calculated value of the entanglement temperature, T$_E$, using the equation \ref{eqn:enttemp} is 390 K implying that at this temperature half of the total dimer spins should have been in the triplet state reducing the total entanglement of copper acetate to zero. The presence of finite entanglement even at temperatures equal to the energy gap of the dimer implies that there is some other energy scale in the system that protects the entanglement of the system against decoherence with the surroundings. The fact that this temperature is much above room temperature is quite surprising! To our knowledge, such a high temperature for entanglement existence in a many-body system has not been previously reported. From the magnetic susceptibility measurements, the exchange coupling constant, J$/k_B$, for the spin dimer has been found to be 430 K. In a recent work, we have shown that multipartite entanglement in a spin 1/2 uniform-chain system \cite{george} exists at temperatures as high as 2.5J$/k_B$. So, it is quite probable that a finite bipartite entanglement exists till at least 430 K in this system as well. Unfortunately, the copper acetate crystal degrades at temperatures larger than 380 K (refer to the TGA discussions above), so we cannot investigate the spin entanglement of this excellent dimer system at temperatures larger than 380 K.\\  

\subsubsection{via specific heat}
\begin{figure}[h!]
	\centering
	\includegraphics[width=1\textwidth]{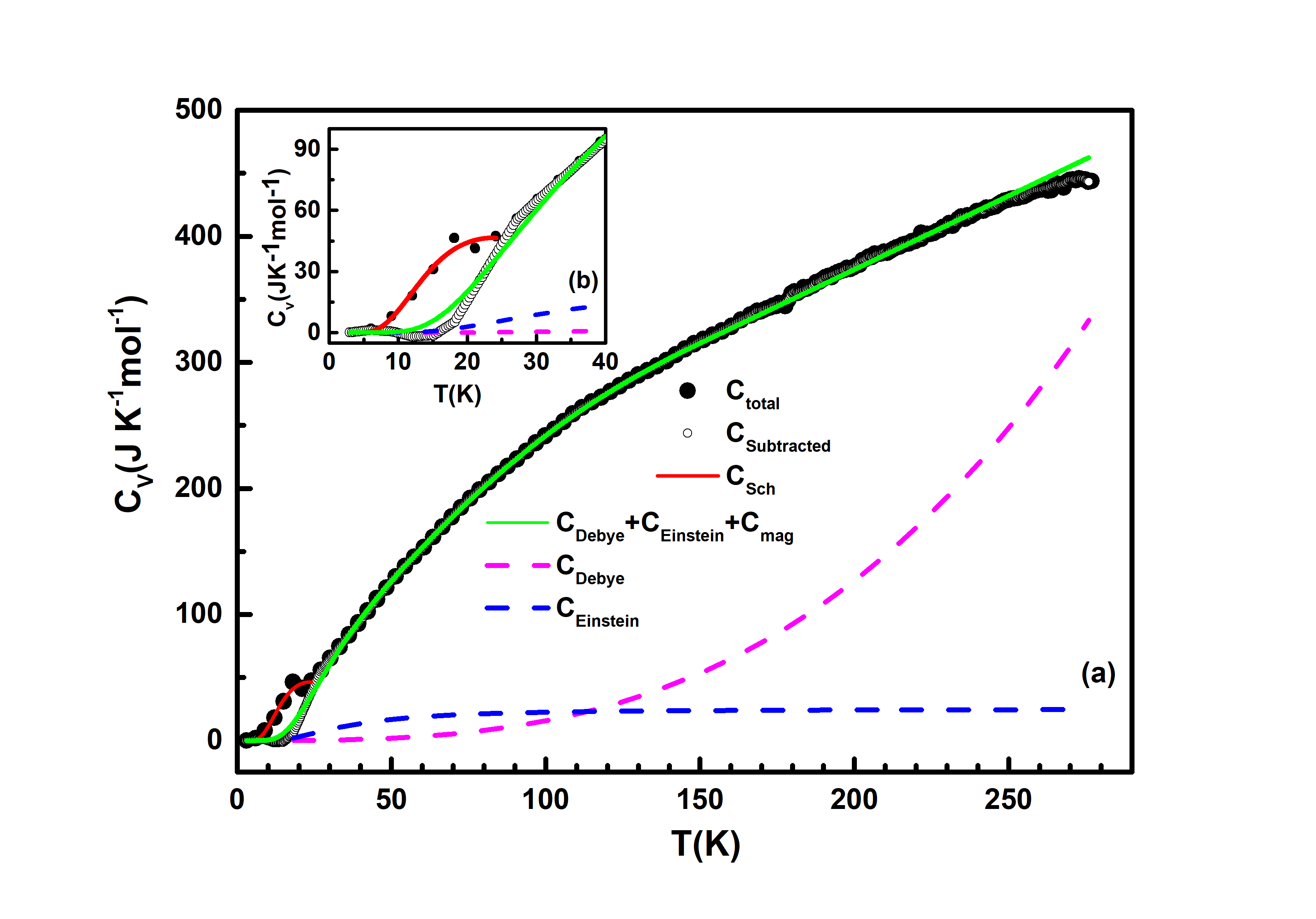}
	\caption{(a) Black filled circles represent the experimentally obtained specific heat data in zero applied magnetic field. Red solid curve is a fit to the Schottky equation \ref{eqn:schottky} while the open black circles represent the data after the subtraction of the Schottky contribution (see text for details). Green solid line represent the fit to equation \ref{eqn:Cv} while pink and blue dotted lines represent simulated data for contributions due to Debye and Einstein respectively. (b) Expanded data in the temperature range 0 to 40 K to show the Schottky peak clearly.}
	\label{fig:Cv}
\end{figure}

In order to compare the distance between the state measure obtained from susceptibility to that obtained from specific heat, we performed specific heat measurements from 3 K to 270 K on copper acetate in zero applied magnetic field as illustrated by the filled black circles in Fig. \ref{fig:Cv}. There are two noteworthy observations from the figure. The first is the observation of a small peak at a temperature of $\sim$ 20 K. Since this temperature is exactly the same as the temperature where the magnetic susceptibility showed an upturn arising due to uncoupled spins (see above), the peak is ascribed to a Schottky anomaly \cite{gopal}. 

To confirm that the peak is indeed due to a Schottky anomaly, we fitted the peak using the Schottky equation \ref{eqn:schottky}, considering a two-level system with S = 1/2, as shown below: 
\begin{equation}\label{eqn:schottky}
C_{Sch}= R\bigg(\frac{\Delta}{k_BT}\bigg)^2\frac{e^\frac{\Delta}{k_BT}}{\big(1+e^\frac{\Delta}{k_BT}\big)^2}
\end{equation}
where $R$ = 8.314 Jmol$^{-1}$K$^{-1}$ is the molar gas constant and $\Delta$ is the energy separation of the two-level system.
The fit using the above equation in the temperature range 3-40 K is shown as a red solid line in Fig. \ref{fig:Cv} and shown on an expanded scale as the inset (a) of Fig. \ref{fig:Cv}. It can be seen that the red solid line fits the peak at 20 K quite well confirming that the peak arises due to Schottky anomaly arising due to impurities.\\
To discount the effect of the Schkottky anomaly arising due to impurities, we subtracted the specific heat contribution, C$_{Sch}$, calculated as above and the resultant curve is plotted as black open circles in Fig. \ref{fig:Cv}. We assume that the total specific heat of copper acetate arises due to two different contributions, one due to lattice phonons and the other due to magnetic contribution, C$_{Mag}$, arising from the thermal population of excited dimer states. To incorporate the effect of lattice phonons, we fitted the temperature variation of C$_{subtracted}$ considering a combined Debye and Einstein model having relative weights of $m$ and $1-m$ respectively. The total phonon and magnetic contribution was, then, fitted with relative weights of $v$ and $1-v$ respectively: 
\begin{equation}\label{eqn:Cv}
C_{subtracted} (T) = v[mC_{Debye}(T)+ (1-m)C_{Einstein}(T)]+ (1-v)C_{Mag}(T)
\end{equation}
where the first and the second term represent the acoustic and optical phonon mode contribution described by Debye and Einstein models respectively \cite{tari}:
\begin{equation}\label{eqn:CDebye}
C_{Debye}(T) = 9R\bigg(\frac{T}{\theta_D}\bigg)^3\int_{0}^{x_D}dx\frac{x^4e^x}{(e^x-1)^2}
\end{equation}
\begin{equation}\label{eqn:CEinstein}
C_{Einstein}(T) = \sum 3R\bigg(\frac{\theta_{Ei}}{T}\bigg)^2\frac{e^{\frac{\theta_{Ei}}{T}}}{(e^{\frac{\theta_{Ei}}{T}-1})^2}
\end{equation}
where $\theta_D$ and $\theta_{Ei}$ represent the Debye and Einstein temperatures respectively. Since a single Einstein term could not fit the entire data, two Einstein terms E$_1$ and E$_2$ were needed to fit the data. The third term in equation \ref{eqn:Cv} is the magnetic contribution to the specific heat and is given by equation \ref{eqn:Cm} above. Green solid line in Fig. \ref{fig:Cv} is the fit to the subtracted data using equation \ref{eqn:Cv}. The best fit was obtained for $\theta_D$ = 487 K, $\theta_{E1}$ = 111 K, $\theta_{E2}$ = 1275 K, $v$ = 0.73 and $m$ = 0.39. The pink and blue dashed lines in the Fig. \ref{fig:Cv} show the simulated data corresponding to Debye and Einstein contribution respectively. To validate the findings of the fit, these temperatures are compared with optical phonon frequencies obtained by Raman spectroscopy on our dimer (refer to Fig. \ref{fig:FTIRRaman} above). The experimentally obtained highest intensity Raman peak at a frequency of 323 cm$^{-1}$ (see Fig. \ref{fig:FTIRRaman} (c)) has an excellent match with the obtained Debye temperature of 495 K which corresponds to a frequency of 345 cm$^{-1}$. Similarly, the Raman peak at 950 cm$^{-1}$ matches fairly well with the obtained Einstein temperature $\theta_{E2}$ = 1275 K (corresponding to 888 cm$^{-1}$). The frequency corresponding to the lower Einstein temperature of $\theta_{E1}$ = 111 K (corresponding to $\sim$ 77 cm$^{-1}$) falls below the measurable limit of the used Raman spectrometer.

\begin{figure}[h!]
	\centering
	\includegraphics[width=1\textwidth]{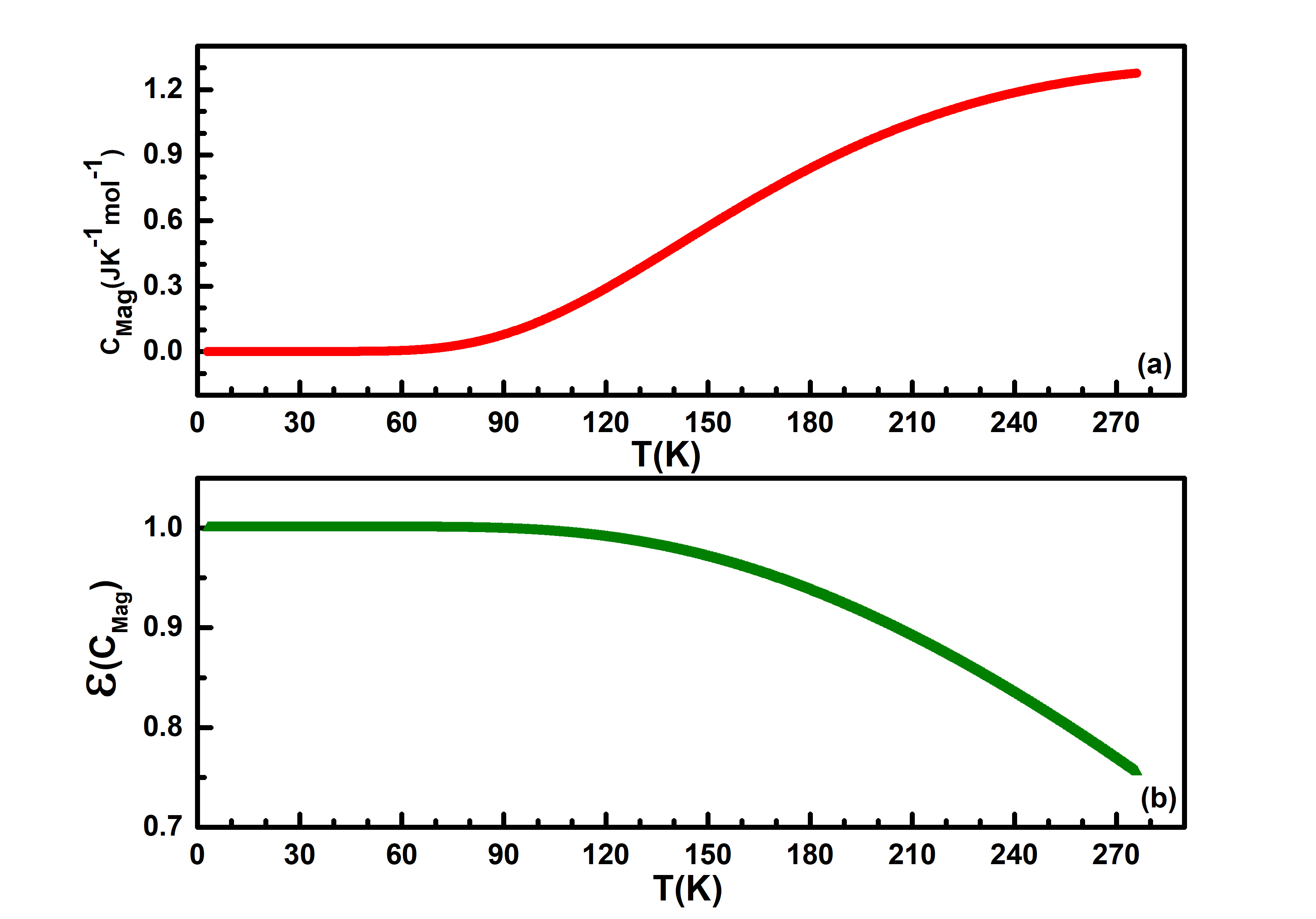}
	\caption{(a) Temperature variation of magnetic contribution to specific heat, C$_{Mag}$, extracted from the fit to equation \ref{eqn:Cv}. (b) Temperature variation of distance between the states in terms of specific heat, $\mathcal{E}(C_{Mag})$, calculated using equation \ref{eqn:entcv} using the extracted specific heat data in (a).}
	\label{fig:ECv}
\end{figure}

To estimate the temperature variation of distance between the states in terms of specific heat, $\mathcal{E}(C_{Mag})$, as given by equation \ref{eqn:entcv}, it is necessary to first plot the temperature variation of magnetic contribution to the specific heat obtained from the magnetic fit to the specific heat, C$_{Mag}$, obtained above. This is plotted as red filled circles in Fig. \ref{fig:ECv} (a). Evidently, the magnetic contribution to the specific heat is quite small compared to the lattice contribution. Since we did not have a non-magnetic reference material using which the lattice contribution could be estimated, we could not unambigously extract the magnetic contribution from the measured specific heat. However, since the temperature variation of C$_{Mag}$ is quite similar to that observed in other published dimer systems \cite{nawa,Eremina}, it is hoped that the temperature variation is of the correct kind. It can be seen that the specific heat value is zero till a temperature of $\sim$ 50 K indicating no density of states at the Fermi level till $\sim$ 50 K since the dimers are in the ground state and the density of triplons is zero. With an increase in temperature, the specific heat increases due to excitation of triplons across the energy gap arising due to the thermal energy. Similar to the observation of a broad peak in magnetic susceptibility at 260 K (see Fig. \ref{fig:susceptibility} (a)), we observe a broad maxima in the specific heat at $\sim$ 260 K, in excellent agreement to the susceptibility data.      

Temperature variation of distance between the states in terms of specific heat $\mathcal{E}(C_{Mag})$, can now be extracted from C$_{Mag}$ obtained in Fig. \ref{fig:ECv} (a), using equation \ref{eqn:entcv} and plotted in Fig. \ref{fig:ECv} (b). From the figure, it can be seen that $\mathcal{E}(C_{Mag})$ is maximum at 1 at the lowest measured temperature of 3 K and stays at 1 till a temperature of $\sim$ 60 K. This behaviour is consistent with the temperature variation of $\mathcal{E}(\chi)$ (see Fig. \ref{fig:entanglement} and discussions therein). Similar to the behaviour of $\mathcal{E}(\chi)$, $\mathcal{E}(C_{Mag})$ starts to decrease above 60 K due to the population of triplons across the energy gap. However, the decrease is a bit slower than $\mathcal{E}(\chi)$. But the two independent calculation shows that finite entanglement exists above room temperature.

\section{Conclusion}
In conclusion, applying a theoretical measure termed as "distance between the states", we have quantified bipartite entanglement of a one dimensional magnetic dimer system via thermodynamical observables like susceptibility and specific heat. Experimental verification of bipartite entanglement was done on copper acetate which is a very well established magnetic dimer and consequently an excellent system to test the bipartite entanglement calculations. Large sized single crystals of copper acetate were grown by a slow evaporation technique and were characterised by techniques such as single crystal XRD, TGA, IR and Raman spectroscopy. In order to investigate the magnetic interactions, we conducted a comparative analysis with DFT computations using dual basis (B3LYP/6-311++G(d,p), B3LYP/LanL2DZ). Delocalisation index was then calculated for different bonds and shown that direct exchange between the copper atoms is not likely and the magnetic interactions occur via indirect superexchange. Temperature variation of both magnetic susceptibility as well as specific heat demonstrates the existence of bipartite entanglement till room temperature demonstrating the potential usefulness of this dimer system for quantum computations.

\begin{acknowledgments}
D. J-N. acknowledges financial support from SERB, DST, Govt. of India (Grant No. CRG/2021/001262). Kakarlamudi Akhil Chakravarthy is acknowledged for his help in calculations during the initial stages of the work.
\end{acknowledgments}

\end{document}